\documentclass[a4paper,10pt,twoside]{cpc-hepnp}
\usepackage{multicol}
\usepackage{graphicx}
\usepackage{booktabs}
\usepackage{amssymb,bm,mathrsfs,bbm,amscd}
\usepackage[tbtags]{amsmath}
\usepackage{lastpage}
\usepackage{multicol}
\usepackage{multirow} 
\usepackage{subfigure}
\usepackage[abs]{overpic}
\usepackage{capt-of}
\usepackage{newcent}
\usepackage{sverb, longtable}
\usepackage{rotating, multirow}
\newcommand{\tabincell}[2]{\begin{tabular}{@{}#1@{}}#2\end{tabular}}
\usepackage{lineno}
\newenvironment{figurehere}{\def\@captype{figure}}{}
\makeatother

\begin{document}

%\linenumbers

\fancyhead[c]{\small Chinese Physics C~~~Vol. xx, No. x (201x) xxxxxx}
\fancyfoot[C]{\small 010201-\thepage}
\footnotetext[0]{Received xx Apr. 2016}

\title{Radiation Studies for the Target Station of the MOMENT\thanks{Supported by National Natural Science Foundation of China (11425524, 11527811, 11575226) and Strategic Priority Research Program of the Chinese Academy of Sciences (Grant No. XDA10010100).}}

\author{%
     Qing-Nian Xu$^{1;1)}$\email{xuqingnian10@mails.ucas.ac.cn}%
\quad Jian-Fei Tong$^{2}$
\quad Nikolaos Vassilopoulos$^{2}$
\quad Jun Cao$^{1}$ \\
\quad Miao He$^2$ 
\quad Zhi-Long Hou$^2$ 
\quad Han-Tao Jing$^2$
\quad Huai-Min Liu$^2$ \\
\quad Xiao-Rui Lyu$^1$
\quad Jing-Yu Tang$^2$%
\quad Ye Yuan$^{2;2)}$ \email{yuany@ihep.ac.cn}
\quad Guang Zhao$^2$ \\
\quad Yang-Heng Zheng$^{1;3)}$ \email{zhengyh@ucas.ac.cn}
}
\maketitle

\address{%
$^1$ University of Chinese Academy of Sciences, Beijing 100049, China \\
$^2$ Institute of High Energy Physics, Chinese Academy of Sciences, Beijing 100049, China\\
}

\begin{abstract}
The discovery of the neutrino mixing angle $\theta_{13}$ opens new opportunities for the discovery of the leptonic CP violation for high intensity neutrino beams. MOMENT a future neutrino facility with a high-power proton beam of 15~MW from a continuous-wave linac is focused on that discovery. The high power of the proton beam causes extreme radiation conditions for the facility and especially for the target station where the pion capture system of five superconducting solenoids is located. In this paper initial studies are performed for the effects of the radiation on the solenoid structure and the area surrounding it. A concept cooling system is also proposed.
\end{abstract}



\footnotetext[0]{\hspace*{-3mm}\raisebox{0.3ex}{$\scriptstyle\copyright$}2013
Chinese Physical Society and the Institute of High Energy Physics
of the Chinese Academy of Sciences and the Institute
of Modern Physics of the Chinese Academy of Sciences and IOP Publishing Ltd}%

\begin{multicols}{2}

\section{Introduction}
\subsection{MOMENT}
In recent years, neutrino physics has a prominent progress. The last neutrino mixing angle 
$\theta_{13}$ is determined to be non-zero by Daya Bay~\cite{daya_bay} in 2012. 
The large value of the $\theta_{13}$ opens the opportunity to discover the 
leptonic CP-violation and detect the neutrino mass hierarchy in the future 
neutrino superbeams~\cite{HK,LBNE,ESS_1}. Leptonic CP violation is a necessary 
ingredient to generate the observed matter dominance in the universe since the 
measured CP violation in the quark sector is not enough to account for 
the matter-antimatter asymmetry in the universe~\cite{cp_1,cp_2}. 

\begin{center}
\includegraphics[width=7cm]{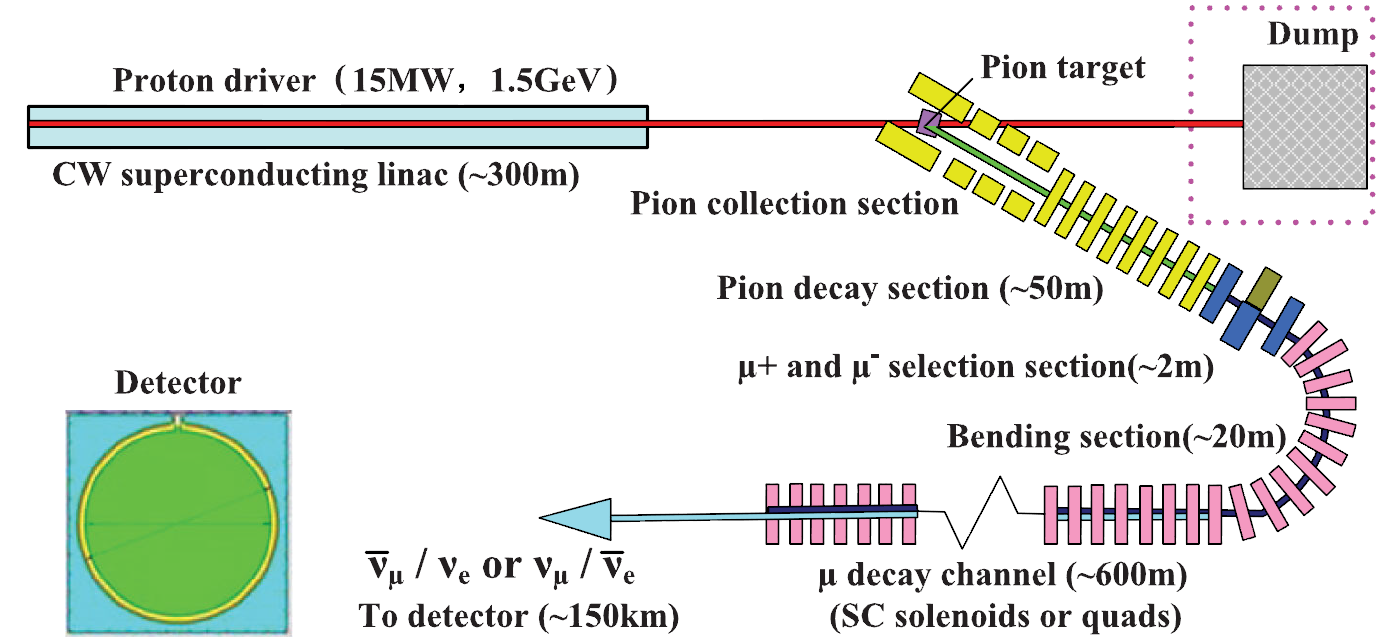}
\figcaption{\label{mom_lay}(color online) Layout of the MOMENT facility\cite{moment}.}
\end{center}

The MOMENT~(MuOn-decay MEdium-baseline NeuTrino beam)~\cite{moment} is a future 
facility focused on the CP phase measurement using neutrinos from muon decays. MOMENT adopts a 15~MW continuous-wave (CW) 1.5~GeV proton beam provided by China-ADS linac~\cite{ch_ads}. Pions are generated from interactions between the proton beam and a liquid mercury jet target immersed in a high magnetic field of a capture superconducting solenoid. High energy protons escaping the interaction region are absorbed by a beam dump near to the target station. After that, a 50~m pion-decay line where pions are decaying to muons and neutrinos is designed. Thereafter a muon charge selection system and a bending transport section in order to separate the direction of the muons and the pion decay neutrinos are being designed. The last part of the beamline consists of an adiabatic transport section and a long muon decay channel of 600 m designed to focus the muons towards the detector direction and allow them to decay to neutrinos respectively. The average neutrino beam energy is $<E_{\nu}>$ = 300~MeV. The detector is foreseen to be located 150~km far away from the facility. The layout of MOMENT experiment is shown in Fig.~\ref{mom_lay}.

\subsection{Target station}
At the target station pions are generated from protons colliding with the mercury target and then collected and transported by adiabatic magnetic fields produced by five superconducting (SC) solenoids. Due to the extremely high radiation large amounts of heat will be deposited on the elements of the solenoids. In addition,  radiation damage will be induced. Therefore, a tungsten cylindrical shield is necessary to protect the five solenoids. The geometry design of the target station then is strongly depended on the limits set from radiation study.

\begin{center}
\includegraphics[width=7cm]{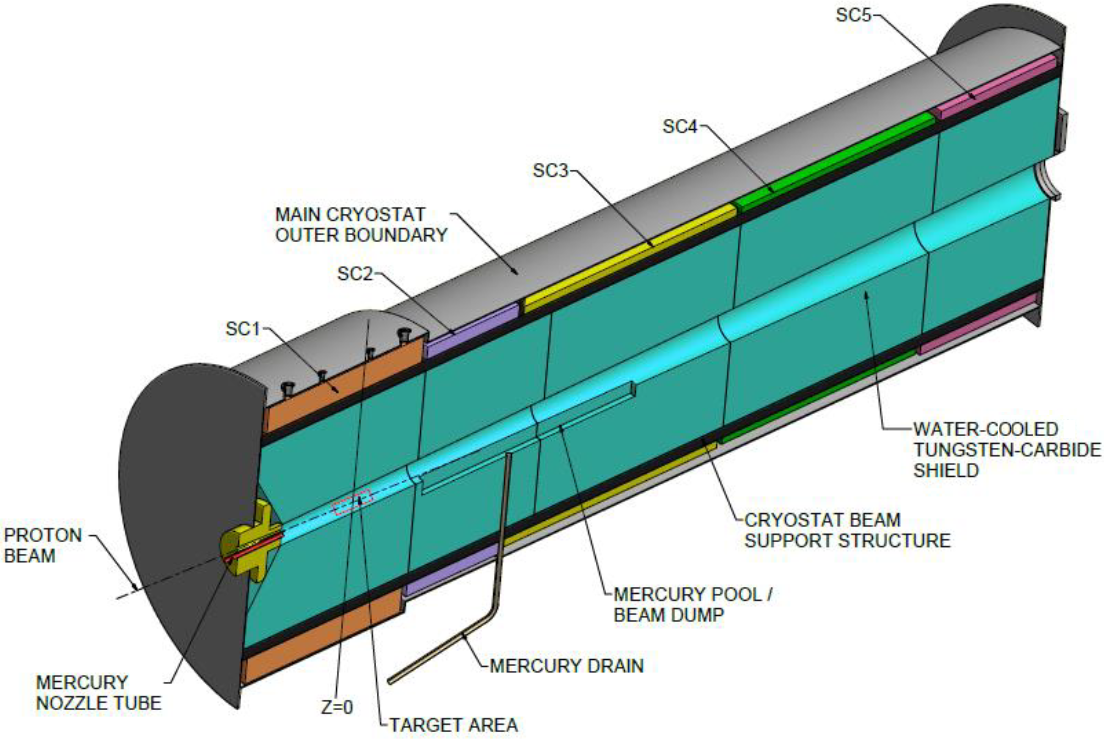}
\figcaption{\label{ts_fig}(color online) Layout of the MOMENT pion capture system solenoid\cite{moment}.}
\end{center}

\begin{center}
\tabcaption{ \label{sc_inf} Geometrical characteristics of the superconducting solenoids. }
\footnotesize
\begin{tabular*}{80mm}{c@{\extracolsep{\fill}}cccccc}
\hline
\hline
\multicolumn{1}{c}{~} & \multicolumn{1}{c}{Material} &  \multicolumn{2}{c}{ Z(m)} 
 &  \multicolumn{2}{c}{ R(m)}  \\ %\cline{2-5}
\hline
             &  & From & To & Inner & Outer  \\
\hline
SC 1         &  \multirow{2}{*}{Nb$_3$Sn} &  -0.7 & 0.99 &  1.05 & 1.23   \\
SC 2         &                            &  1.04 & 2.08 &  1.05 & 1.14   \\
SC 3         &  \multirow{3}{*}{NbTi}     &  2.13 & 4.33 &  1.05 & 1.13   \\
SC 4         &                            &  4.38 & 6.58 &  1.05 & 1.11   \\
SC 5         &                            &  6.63 & 7.63 &  1.05 & 1.14   \\
\bottomrule
\end{tabular*}
\vspace{0mm}
\end{center}
\vspace{0mm}

\begin{center}
\includegraphics[width=7cm]{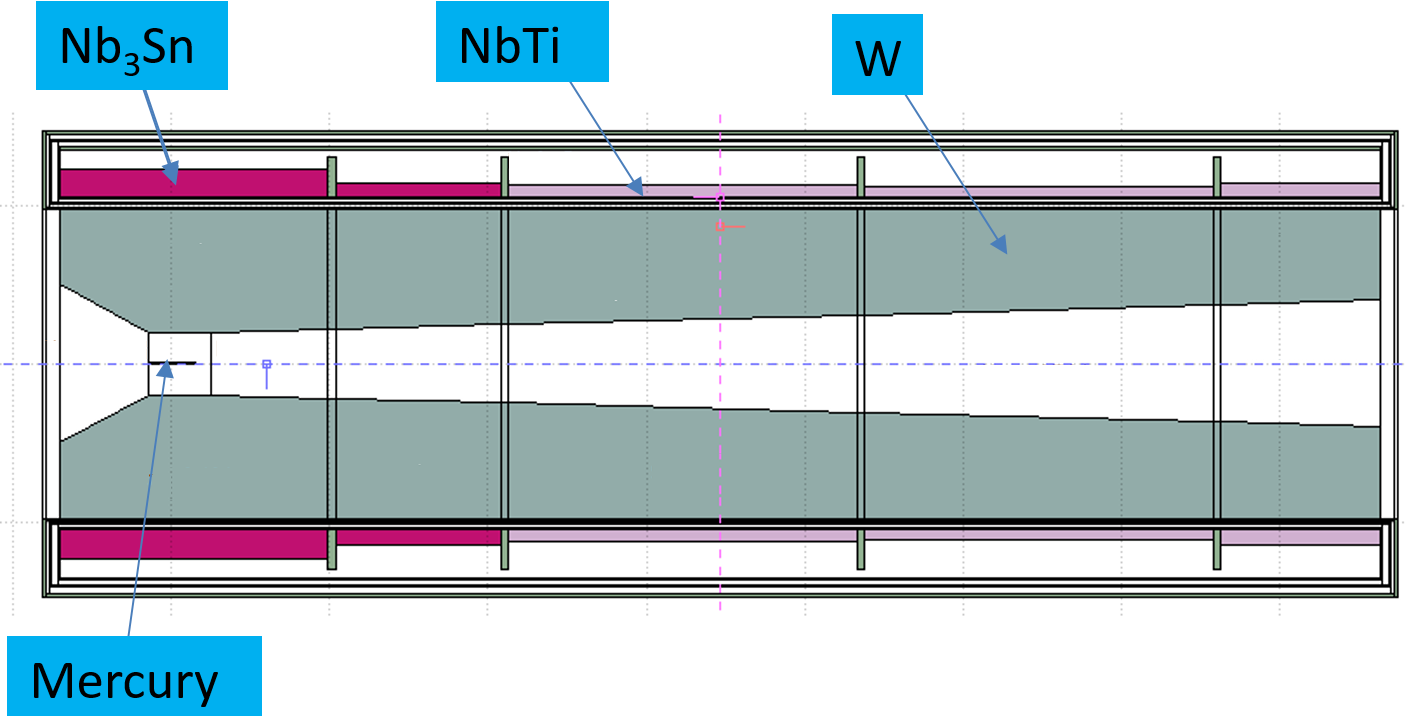}
\figcaption{\label{sc_she}(color online) Transverse view of the tungsten shield and the five superconducting solenoids from FLUKA simulation.}
\end{center}

The target station mainly consists of the target and the solenoid as shown in Fig.~\ref{ts_fig}. Solid targets used in the conventional (low intensity) neutrino beams could not withstand such high beam power so fluid targets as liquid or powder jets and waterfalls are being studied. Their main advantages are minimum material damage, and high heat absorption and transfer due to their recycling nature. 
At MOMENT the nominal mercury jet target is placed in a high-field superconducting solenoid~\cite{merit}. A 14~T field is used to capture the charged mesons with high efficiency and then a slow adiabatic decrease is implemented from 14~T to 3-4~T in order to reduce their transverse momentum (with respect to the beam-line direction) and maximize their transport efficiency. In order to produce these fields, five superconducting solenoids are used with a total length of about 8.3~m and a radius of about 1~m made by Nb$_3$Sn and NbTi wires as summarized in Table~\ref{sc_inf}. A tungsten shield with a maximum (minimum) thickness of 77.5 (57.8)~cm at the beginning (end) is placed between the target and the solenoid. This solenoid configuration is proposed in \cite{moment} and is studied in this paper. FLUKA Monte Carlo~\cite{fluka_1,fluka_2}  is used to study the energy deposition, radiation damage and activity of the tungsten shield and the superconducting solenoids.

%FLUKA introduction 
\section{Monte Carlo simulation with FLUKA}
FLUKA Monte-Carco is used to simulate hadronic and electromagnetic interactions that take place in several experiments. It is one of the main codes used for radiation studies and safety calculations as energy depositions and radiation damage in materials along with particle fluxes and dose rates among others.  It is continuously validated with data from low energy nuclear physics, high energy experiments and atmospheric fluxes. FLUKA uses a resonance model to simulate hadron-nucleon interactions below few GeV whereas the Dual-Parton Model (DPM) used above them. The hadron-nucleus interactions are treated with the PreEquilibrium Approach to NUclear Thermalization model, including the Gribov-Glauber multi-collision mechanism followed by the pre-equilibrium stage and eventually equilibrium processes (evaporation, fission, Fermi break-up and gamma deexcitation). FLUKA can simulate with high accuracy the interaction and propagation in matter of particles from 1 keV to thousands of TeV and especially neutrons down to thermal energies \cite{fluka_MC}.
For the calculation of the the $dpa$ in material, FLUKA uses an equivalent partition function to the Lindhard one applied in other codes, with reworked formulas and restrictions in energy above a user defined damage threshold \cite{fluka_dpa}.The energy deposition, radiation damage and activity is calculated with mesh of Z:R:$\phi$ = 1~cm~:~1~cm~: 2$\pi$ and biasing method is not used in the calculation \cite{fluka_manue}.

\begin{center}
\includegraphics[width=7cm]{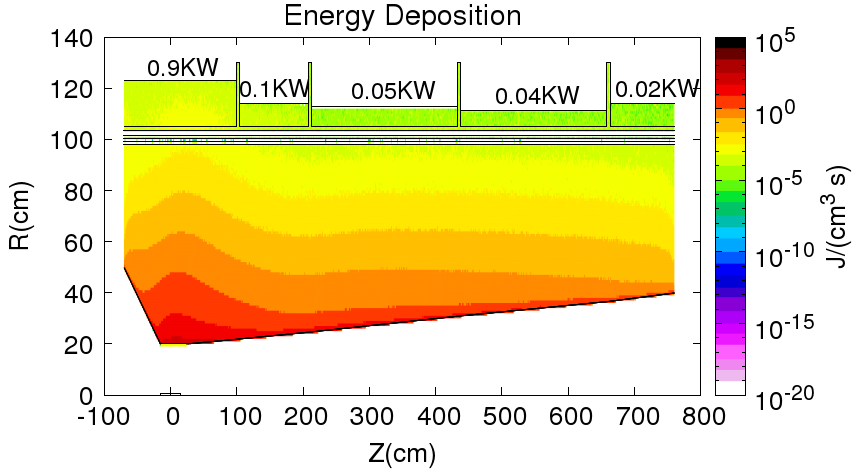}
\figcaption{\label{endp}(color online) Energy deposition density as function of depth for the shield and superconducting solenoids.}
\end{center}

\section{Energy deposition}
The energy deposited on the shield and the solenoid is calculated. The primary mechanism of the energy deposition is the ionization of the atoms by heavy charged particles, electromagnetic showers by electrons and gammas~\cite{leo_book} and nuclear interactions of neutrons. The shield is needed to protect the superconducting solenoids (Nb$_3$Sn, NbTi \cite{nb3sn_sc}) from heating and structural damage. The transverse view of the simulated geometry in FLUKA is shown in Fig.~\ref{sc_she}. The energy deposition density on the materials is shown in Fig.~\ref{endp}. The total energy deposition on the shield is about 10~MW and the maximum volumetric heat is above 100~W/cm$^3$ around the mercury target area and along the inner part of the shield. The energy deposition in superconducting solenoids is limited below 1~kW, which is acceptable for the cryogenic system of the coils.

\begin{center}
\includegraphics[width=7cm]{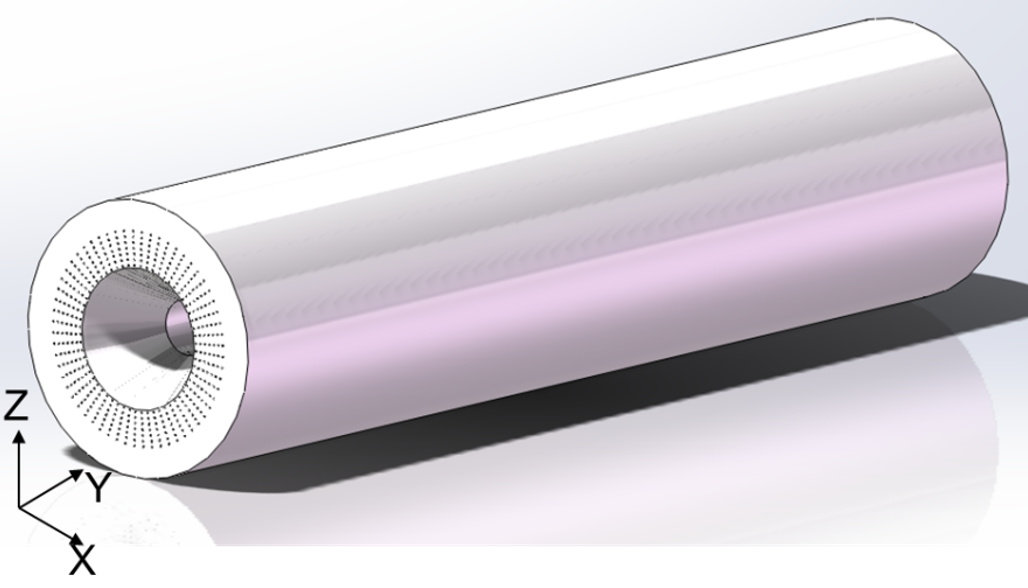}
\put(-120,10){(a)}

\includegraphics[width=7cm]{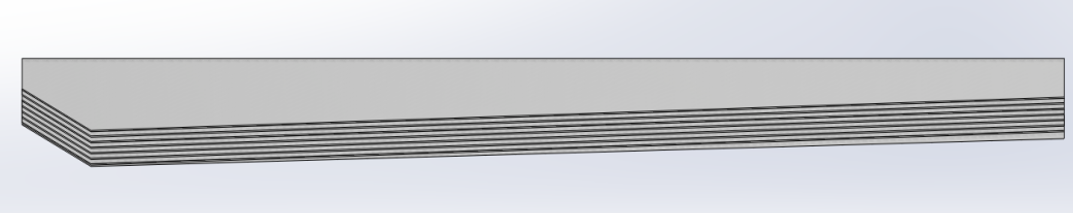}
\put(-120,0){(b)}

\figcaption{\label{cool}(color online) Geometry of shield with cooling channel; (a) 3d view of shield; (b) cut view of x=0.}
\end{center}

\section{Cooling structure}
A Multiple Rows of Mini-Channel (MRMC) cooling structure is designed for the tungsten shield as shown in Fig.~\ref{cool}. The size of the cooling channel is 1 cm$\times$1 cm and the first wall thickness (the distance of the channel away from the inner wall of shield) is about 1 cm. The shape of the channel in the shield was designed for removing the highest volumetric heat. The volume ratio of cooling channels is 1\% compared to the total volume of shield.

\begin{center}
\includegraphics[width=7cm,height=4.cm]{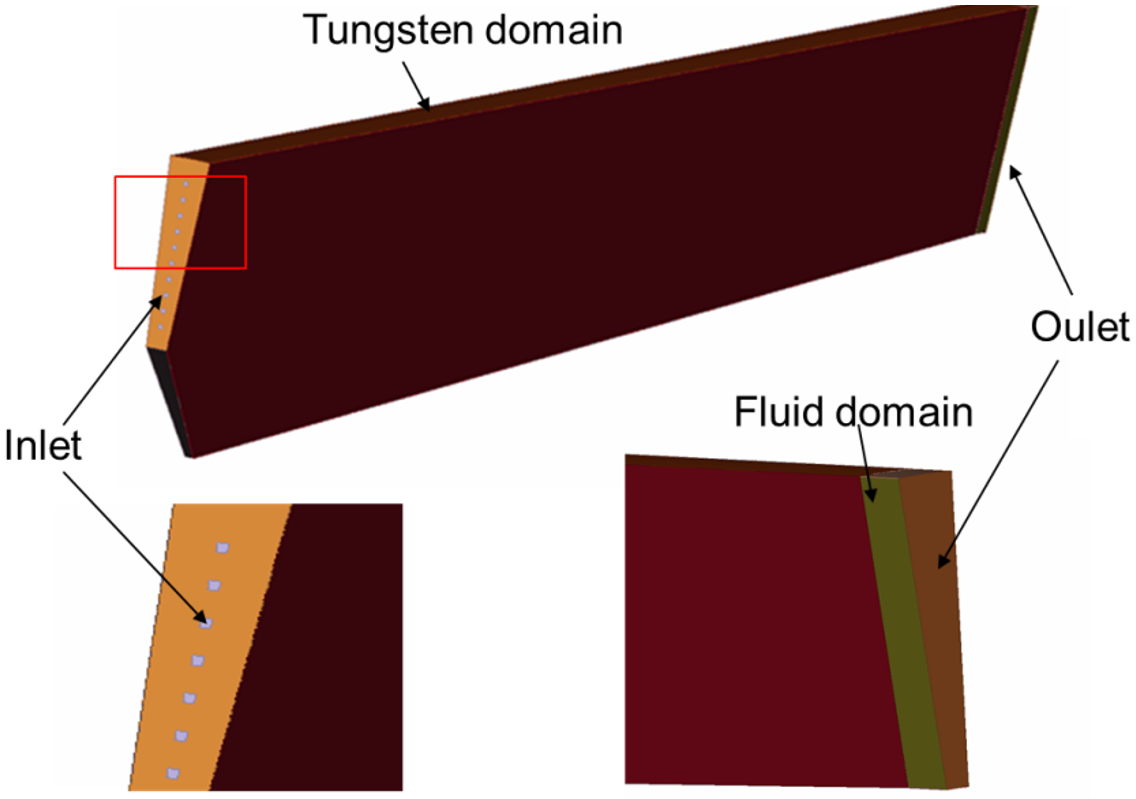}     
\figcaption{(color online) Calculation domain for the shield.}
\label{tung_cool}
\end{center}

\begin{center}
\tabcaption{ \label{the_par} Thermal Parameters of materials. }
\footnotesize
\begin{tabular*}{80mm}{c@{\extracolsep{\fill}}ccccc}
\hline
\hline
  & \tabincell{c}{K\\(W/m$\cdot$ K)} &  \tabincell{c}{Cp\\(J/Kg$\cdot$ K)} &  \tabincell{c}{Viscosity\\(Pas)} & \tabincell{c}{Density\\(Kg/m$^3$)}  \\ %\cline{2-5}
\hline
Water                  & 0.61 &  4181.7  & 8.90$\times$10$^{-4}$ & 9.97$\times$10$^{2}$ \\
\tabincell{c}{Helium@\\30 atm\&\\300K}    & 0.16  &  5181.0  & 2.01$\times$10$^{-5}$ & 4.78 \\
Tungsten               & 1.2$\times$10$^{2}$    &   132.0 &  & 1.94$\times$10$^{4}$\\
\bottomrule
\end{tabular*}
\vspace{0mm}
\end{center}
\vspace{0mm}

A three-dimensional simulation of thermal flow conjugated with solid heat transfer in ANSYS CFX for the  shield is carried out, by compiling a corresponding FORTRAN program in order to describe the non-uniform heat source distribution. The standard $k-\epsilon$ model is used for turbulent dissipation in this calculation. The model is shown in Fig.~\ref{tung_cool}. The mesh number of the solid and the fluid domains are 1.3 million and 0.6 million respectively with the grid dependence checked. Water at 3 atm or helium at 30 atm are examined as possible cooling solutions in this paper. The inlet temperature is 300 K and the velocity of the coolants in each inlet for each channel is 5 m/s and 50 m/s for water and helium respectively. The thermal parameters of materials are shown in Table~\ref{the_par}.

\begin{center}
\includegraphics[width=7cm,height=1.7cm]{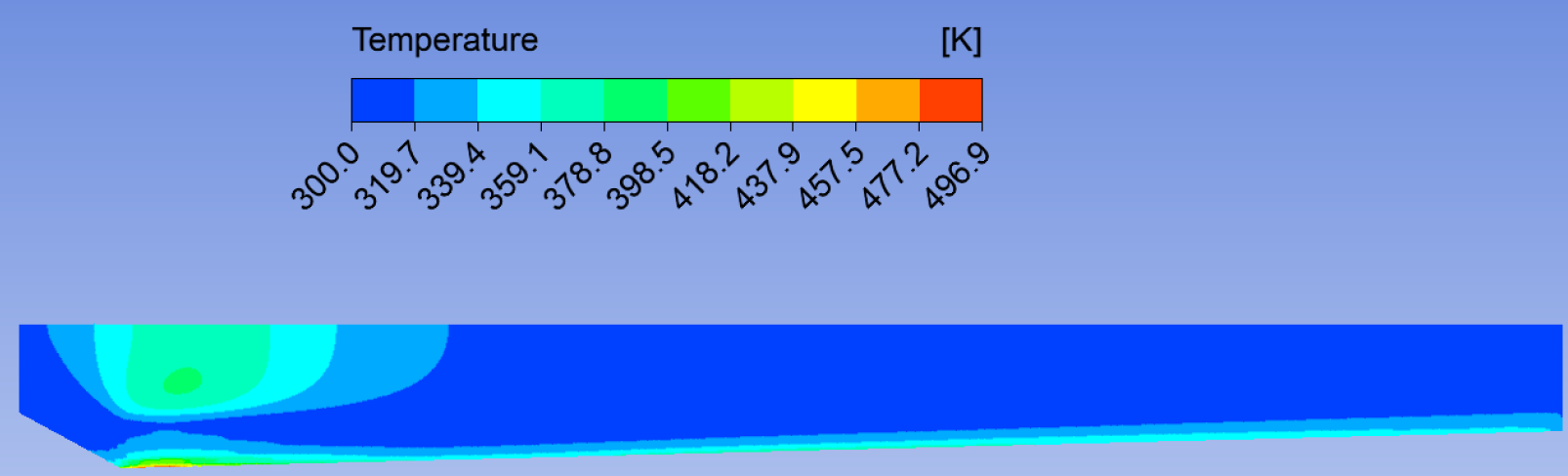}     
\figcaption{(color online) Temperature of the shield with water cooling.}
\label{water_cool}
\end{center}

\begin{center}
\includegraphics[width=7cm,height=1.7cm]{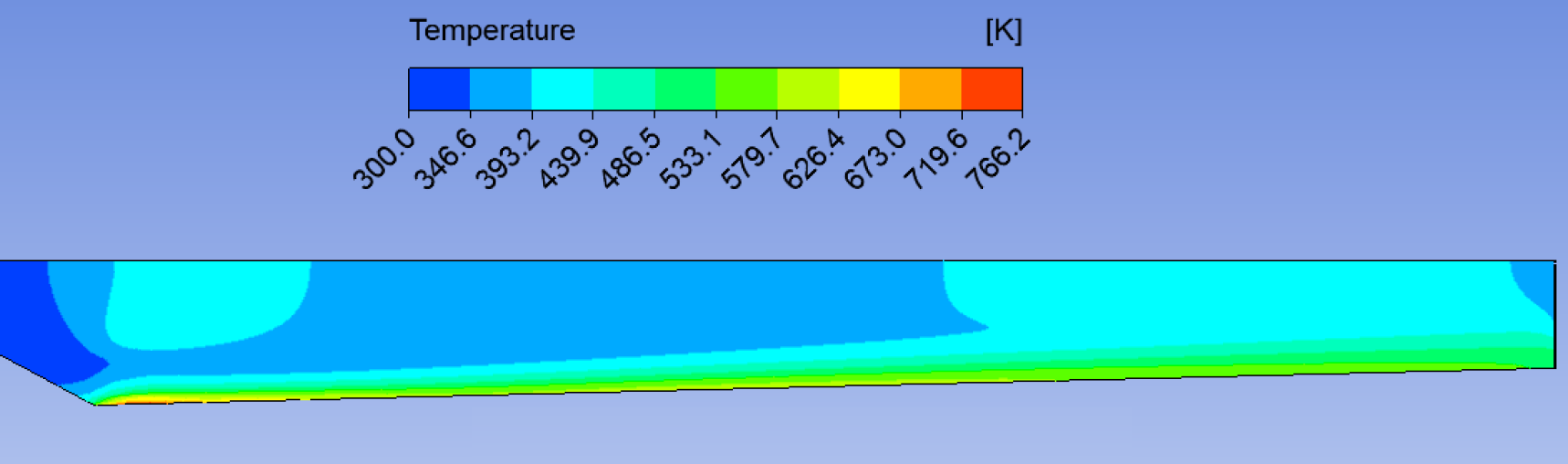}
\figcaption{(color online) Temperature of the shield with pressured helium cooling.}
\label{helium_cool}
\end{center}

\begin{figure*}[!htb]
\centering
\subfigure[]{
\includegraphics[width=5cm,height=3cm]{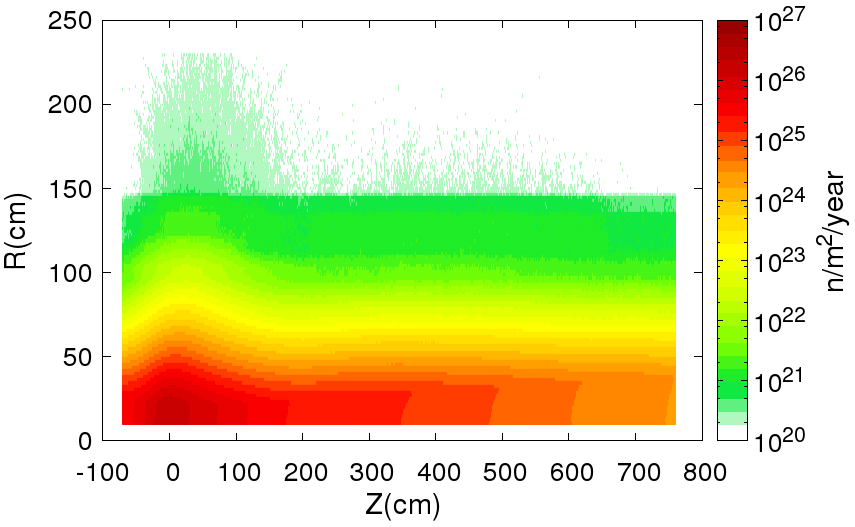}     
\put(-75,85){$n$}
}
\subfigure[]{
\includegraphics[width=5cm,height=3cm]{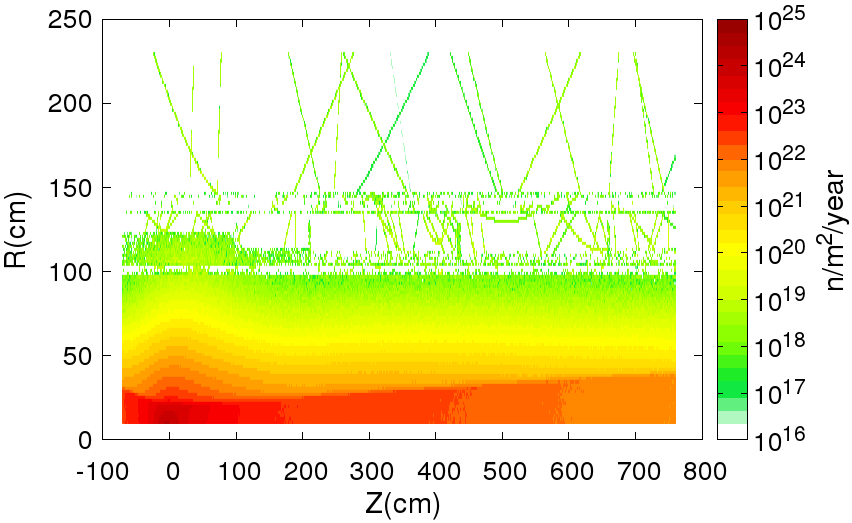}
\put(-75,85){$e^-$}
}
\subfigure[]{
\includegraphics[width=5cm,height=3cm]{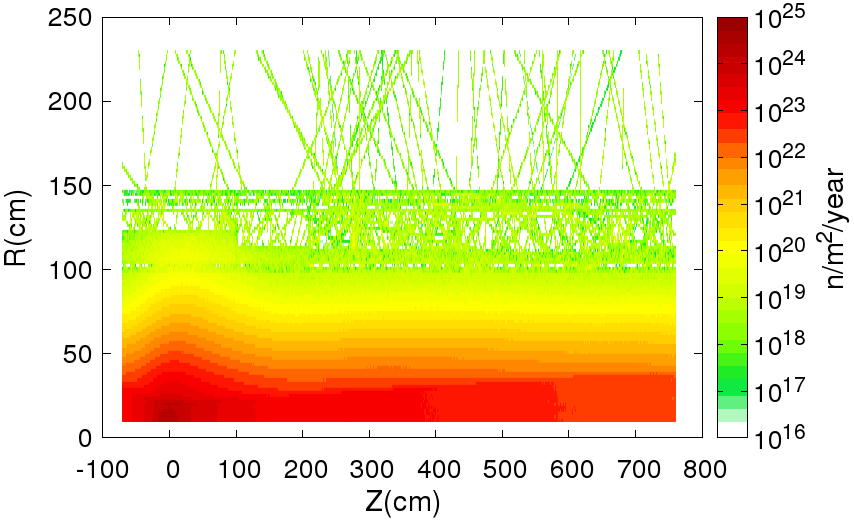}
\put(-75,85){$e^+$}
}

\subfigure[]{
\includegraphics[width=5cm,height=3cm]{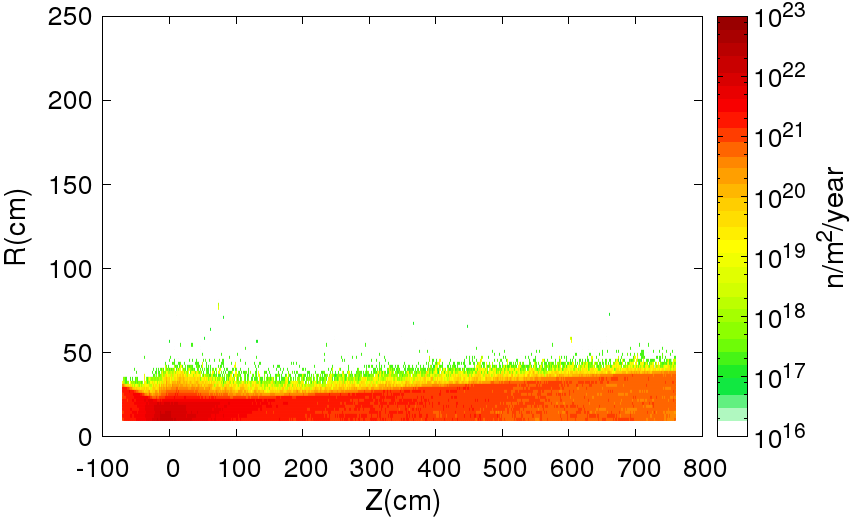}     
\put(-75,85){$\mu^{\pm}$}
}
\subfigure[]{
\includegraphics[width=5cm,height=3cm]{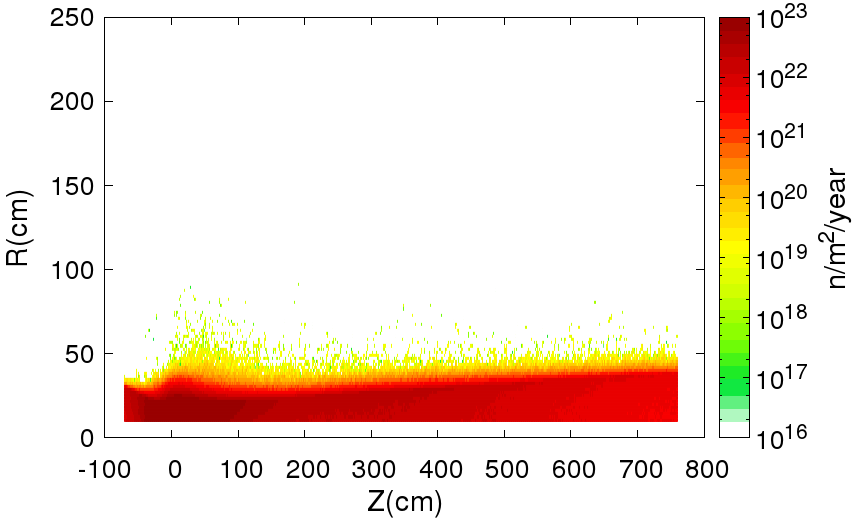}
\put(-75,85){$\pi^{\pm}$}
}
\subfigure[]{
\includegraphics[width=5cm,height=3cm]{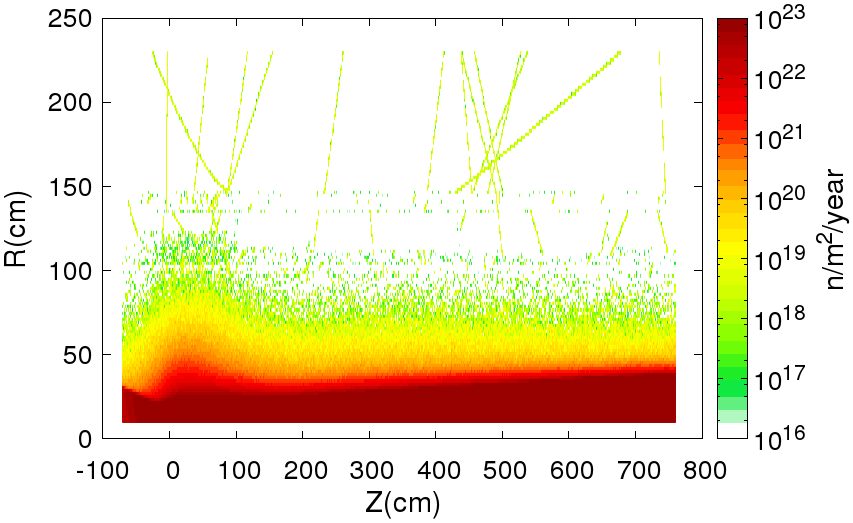}
\put(-75,85){$p$}
}
\caption{(color online) Neutron(a), electron(b), positron(c), muon(d), pion(e) and proton(f) flux distributions as function of depth in the target station.}
\label{neuflx}
\end{figure*}

With the MRMC structure and water as coolant, when the mass flow rate is 3.49 kg/s for this domain, the maximum temperature of the shield is less than 500 K, and the maximum temperature of water is 384 K. The pressure drop is 0.8 MPa and the outlet temperature is 311 K. The results of water cooling are shown in Fig.~\ref{water_cool}. With high pressured helium at high velocity, when the mass flow rate is 0.15 kg/s for this domain, the maximum temperature of the shield is below 900 K, the pressure drop is 0.54 MPa, and the outlet temperature is 519.3 K. The results of helium cooling are shown in Fig. \ref{helium_cool}. The results show that with the MRMC structure and water or pressured helium as coolants, the maximum temperatures for the shield are below 800 degrees, and thus meeting the tungsten cooling demands although with high coolants pressures drops.

\begin{center}
\includegraphics[width=7cm]{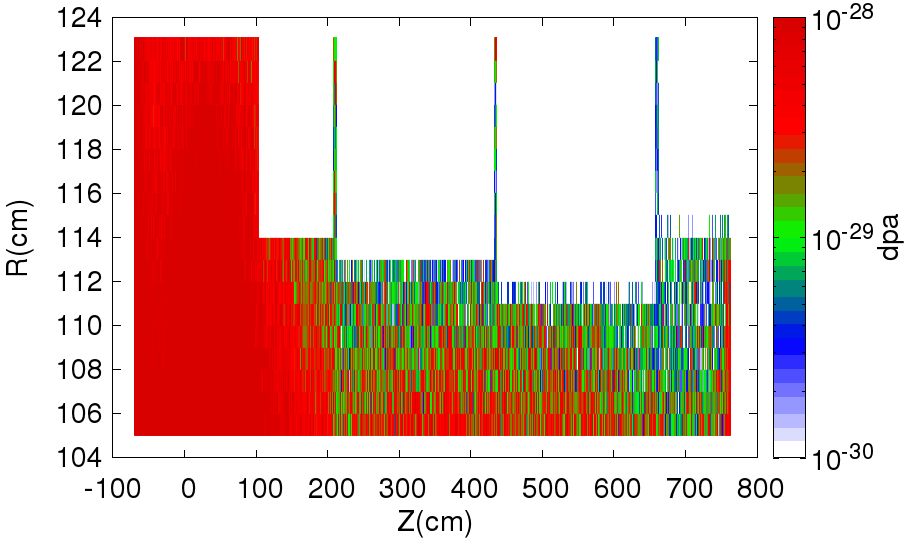}
\figcaption{\label{dpa_sole}(color online) Displacement damage per proton on target. The maximum $dpa$ is calculated around the mercury target and the first SC solenoid. Then it is reduced as function of the length of the solenoid.}
\end{center}

\section{Radiation damage}
Under the 15 MW proton beam interactions with the mercury target, severe radiation is taking place in the space surrounding the target and especially in the different parts of the solenoid. The radiation induced damage for the shield and the superconducting solenoids is calculated in terms of the average displacements of each atom ($dpa$), which are caused by neutrons, charged particles and high energy photons, and is calculated by FLUKA Monte Carlo. The neutron and charged particles flux are shown in Fig.~\ref{neuflx}. It is estimated to have a maximum value of neutron flux about 2$\times 10^{22}$  and 2$\times 10^{21}$ n/m$^2$/year for the first two superconducting solenoids (Nb$_3$Sn) and for the rest three (NbTi) respectively. Damage to Nb-based superconductors appears to become significant at doses of 2-3$\times 10^{22}$ n/m$^2$ \cite{flux_ref}. Reviews of these considerations for ITER \cite{flux_iter}.

\begin{center}
\includegraphics[width=7cm,height=4cm]{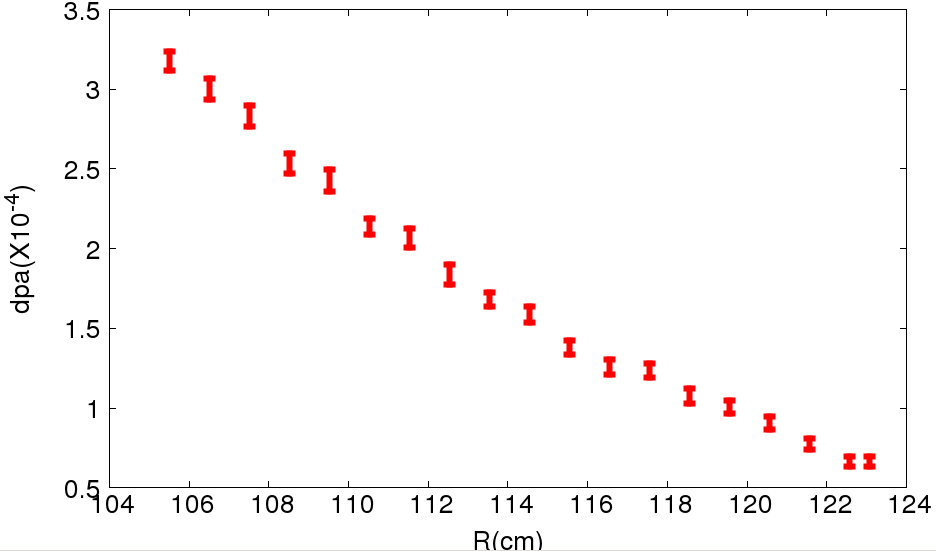}     
\figcaption{\label{dpa_sole1}(color online) Displacement damage dependence as function of depth for the first  superconducting solenoid. The $dpa$ results  normalized by the number of protons for 1~year.}
\end{center}
 
\begin{figure*}[!ht]
\centering
\subfigure[]{
\includegraphics[width=7cm,height=4cm]{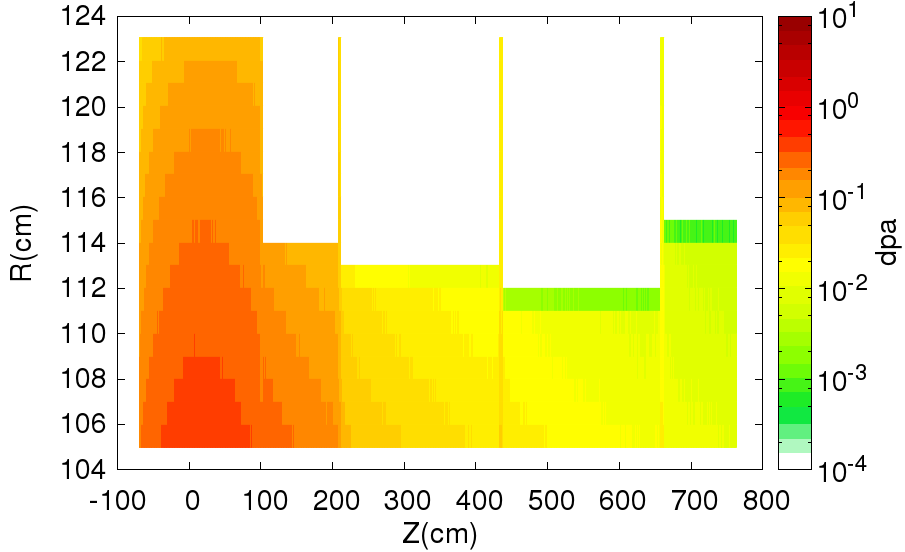}     
\put(-130,80){SC solenoid}
}
\subfigure[]{
\includegraphics[width=7cm,height=4cm]{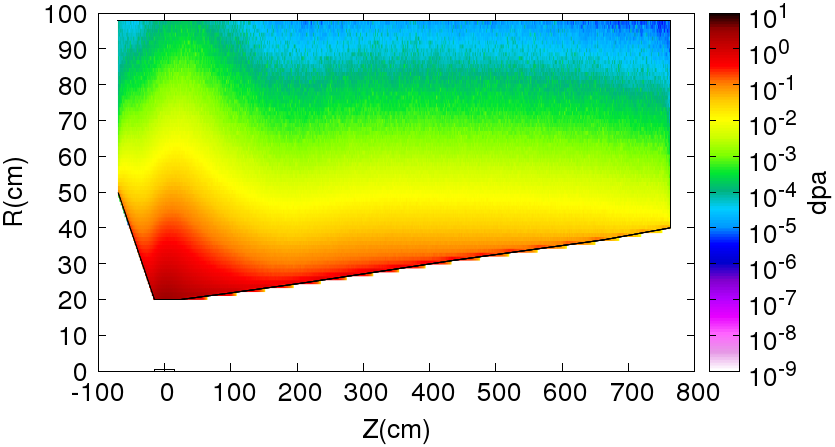}
\put(-120,80){Shield}
}
\caption{(color online) Displacement damage dependence as function of depth for the SC solenoid without shield protection(a) and the shield(b). The $dpa$ results  normalized by the number of protons for 1~year.}
\label{dpa_shie1}
\end{figure*}

\begin{figure*}[!htb]
\centering
\subfigure[]{
\includegraphics[width=7cm,height=4cm]{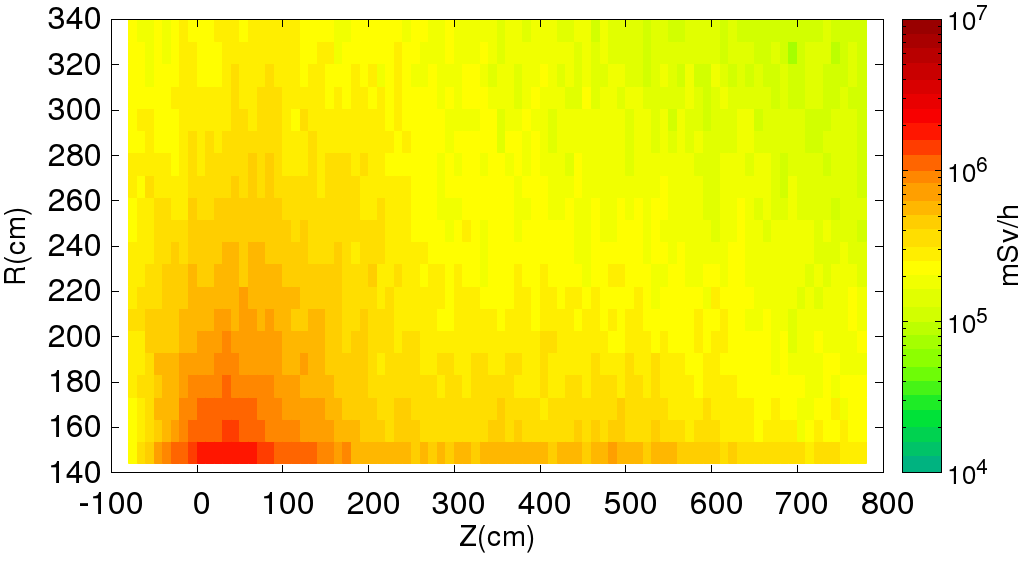}     
\put(-140,85){Prompt dose rate}
\put(-140,70){above SC solenoids}
}
\subfigure[]{
\includegraphics[width=7cm,height=4cm]{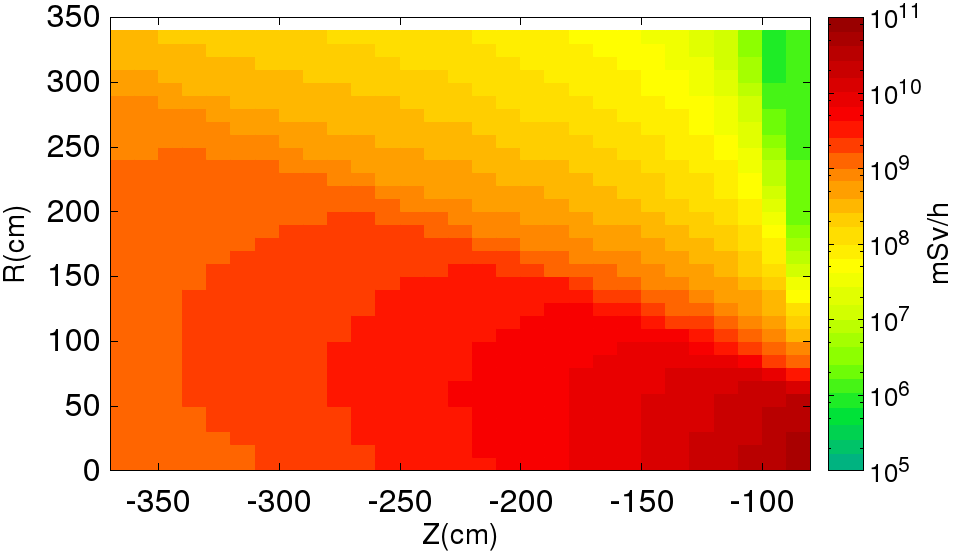}     
\put(-140,85){Prompt dose rate}
\put(-140,70){upstream}
}

\subfigure[]{
\includegraphics[width=7cm,height=4cm]{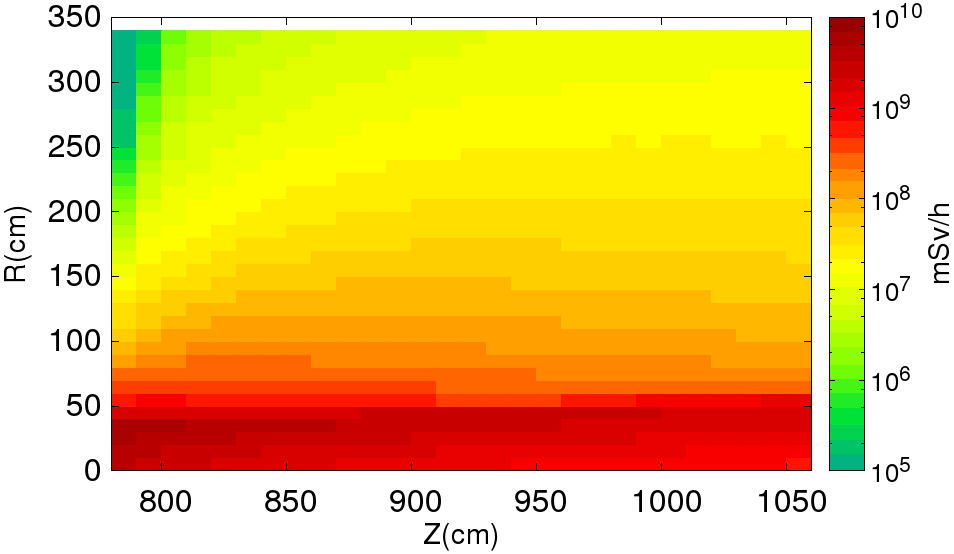}     
\put(-140,85){Prompt dose rate}
\put(-140,70){downstream}
}
\subfigure[]{
\includegraphics[width=7cm,height=4cm]{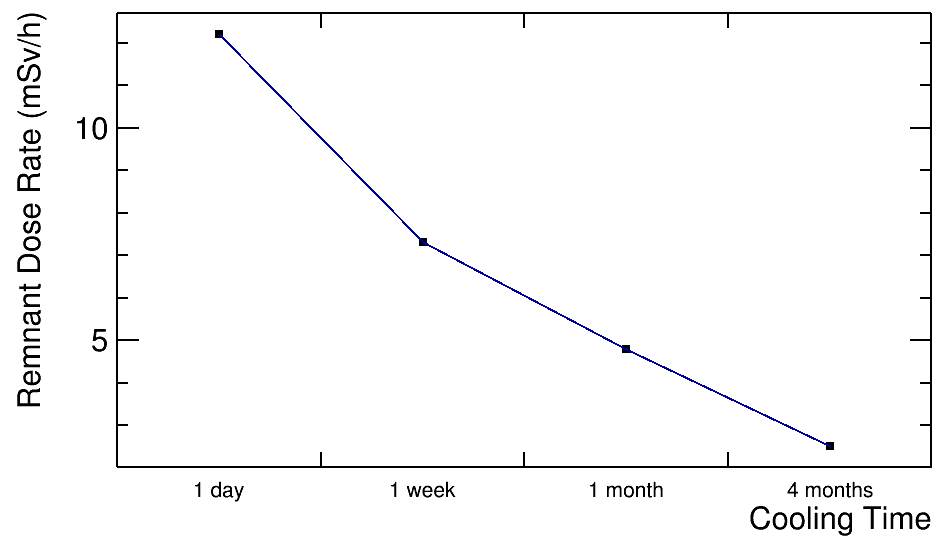}
\put(-120,85){Remnant dose rate}
}
\caption{(color online) Prompt(a),(b),(c) and remnant (d) effective dose rates in the space beyond the SC solenoids is calculated with the EWT74 fluence-to-effective dose conversion coefficients. For the remnant dose rates the value is for a small volume above and in the middle for four cooling times after one year of operation.}
\label{pdos}
\end{figure*}

The $dpa$ per proton on target for the five superconducting solenoids is shown in Fig.~\ref{dpa_sole}. It can be noticed that the largest values are calculated around the mercury target and in the inner locations. The distribution of $dpa$ as function of radius of the first superconducting solenoid for one year of operation~\footnote{For a year, 208 operational days for the accelerator are taken and for a proton beam of 15~MW and 1.5~GeV kinetic energy 1.1$\times$10$^{24}$~p.o.t. are expected.} is shown in Fig.~\ref{dpa_sole1}. The maximum value is about 3.2$\times 10^{-4}$~$dpa$ after one year running. 
The superconductor solenoids mainly consists of the superconductor wire(Nb3Sn/NbTi) and aluminum layer. The damage for aluminum can recover by thermal cycling to room temperature\cite{alum_recov}.
The situation for the superconducting solenoids without shield protection and the 
tungsten shield is totally different. The results are shown  for one year running in Fig.~\ref{dpa_shie1}. 
The values of 0.40 $dpa$ and  2.46~$dpa$ for superconducting solenoids and tungsten shield 
are of concern. These are high value and the $dpa$ for shield  is compared for example with the results obtained in Los Alamos National Laboratory~\cite{dpa_W}. It indicates deterioration of inner part of the tungsten shield and its replacement periodically.

\begin{figurehere}
\centering
\includegraphics[width=7cm,height=4cm]{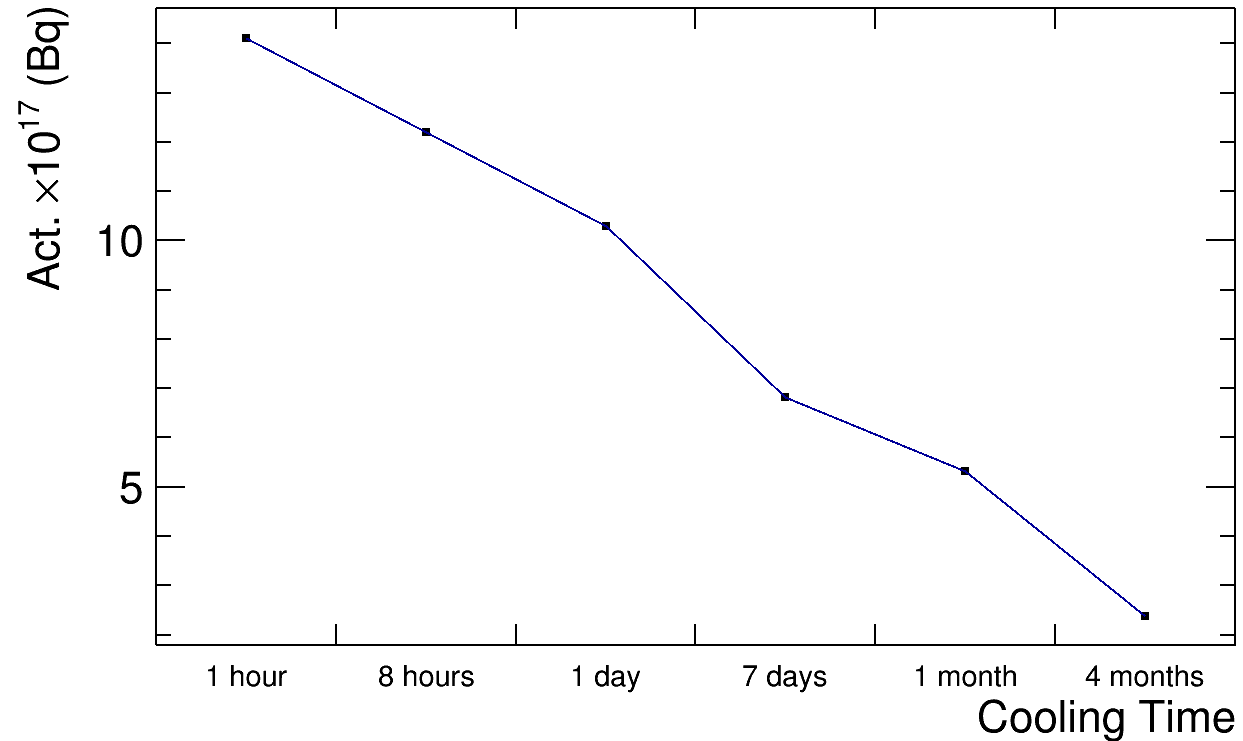}
%\figcaption{\label{activation} Activation for the shield for six cooling times after one year of radiation. The max value reaches 14$\times 10^{17}$ Bq.}
\figcaption{\label{activation} Activity of the shield for six cooling times after one year of radiation. The max value reaches 14$\times 10^{17}$ Bq.}
%\includegraphics[width=.9\textwdith]{activation}
%\caption{ Activation for the shield for six cooling times after one year of radiation. The max value reaches 14$\times 10^{17}$ Bq.}
\label{activation}
\end{figurehere}

\section{Study of effective dose rate and activity}
The prompt and remnant effective dose rates are calculated in the surrounding areas of the solenoid in order to examine their radiation levels. The prompt dose rate reaches $8.9\times10^6$, $4.2\times10^{10}$ and $5.6\times10^9$~mSv/h in the areas above, upstream and downstream of the solenoid respectively as shown in Fig.~\ref{pdos}. For the floor above the solenoid, future studies are needed to define the depth and the composition of the shield  in order to lower the dose rates to acceptable limits at $\mu$Sv/h levels according to the radiation protection rules~\cite{website}. The remnant dose rate is also calculated after one year of irradiation and for a day, a week, a month and four months cooling times. They are measured at a small volume above the center of the superconducting solenoids and are shown in Fig.~\ref{pdos}. The value of the remnant dose rates as function of cooling time is decreasing as expected but still the values are high after four months, of the order of~mSv/h.
It's dangerous for the maintenance worker of this radiation exposure , a remote maintenance is advised.
With the addition of the shield surrounding the experimental layout these values would be expected to decrease at radiation protection limits. Finally, the total activity of the shield is shown in Fig.~\ref{activation}.
This activity is still preliminary and more work should be carried out. It could be done in the future.

\section{Conclusion}
MOMENT uses a high-power beam of 15 MW from a CW linac, which is a challenge for any target station and collection scheme due to the high  radiation. A sophisticated shielding  is necessary to protect the superconducting solenoids. The calculation of the energy deposition in the shield, superconductors and cryostat has been carried out, and a cooling structure for the inner shield by using either water or high-pressure helium is also studied. The problem arises from the radiation damage of the shield. The displacements per atom calculations indicate that fractures will be created on the shield thus periodical replacement is necessary. The low values of the radiation damage for the solenoid wire materials indicate that the shield protection is highly effective. The prompt and remnant effective dose rates are also calculated above the solenoid. As result heavy shielding surrounding the solenoids has to be studied in order to keep the dose rates values according to the radiation production limits. 

\end{multicols}

\begin{multicols}{2}
\end{multicols}

\vspace{15mm}

\vspace{-1mm}
\centerline{\rule{80mm}{0.1pt}}
\vspace{2mm}

\begin{multicols}{2}

\end{multicols}

\clearpage
\end{document}